\documentclass[prb,lengthcheck]{revtex4-1}

\usepackage{amsmath}
\usepackage{bm}
\usepackage{amssymb}
\usepackage{graphicx,color}
\usepackage{hyperref}
\usepackage{braket}
\usepackage{textcomp}

\newcommand{\operator}[1]{\ensuremath{\hat{#1}}}
\newcommand{\ic}{\ensuremath{\mathrm{i}}}
\newcommand{\rket}[1]{\ensuremath{|#1)}}
\newcommand{\rbra}[1]{\ensuremath{(#1|}}
\newcommand{\rbraket}[1]{\ensuremath{(#1)}}

\begin{document}

\title{Variational matrix product ansatz for dispersion relations}

\author{Jutho Haegeman$^{1}$}
\author{Bogdan Pirvu$^{2}$}
\author{David J.\ Weir$^{3}$}
\author{J.\ Ignacio Cirac$^{4}$}
\author{Tobias J.\ Osborne$^{5}$}
\author{Henri Verschelde$^{1}$}
\author{Frank Verstraete$^{2,6}$}
\affiliation{$^1$Ghent University, Department of Physics and Astronomy, Krijgslaan 281-S9, B-9000 Ghent, Belgium\\
$^2$University of Vienna, Faculty of Physics, Boltzmanngasse 5, A-1090 Wien, Austria\\
$^3$Theoretical Physics Group, Blackett Laboratory, Imperial College London, London SW7 2AZ, U.K.\\
$^4$Max-Planck-Institut f\"ur Quantenoptik, Hans-Kopfermann-Str. 1, Garching, D-85748, Germany\\
$^5$Leibniz Universit\"at Hannover, Institute of Theoretical Physics, Appelstrasse 2, D-30167 Hannover, Germany\\
$^6$C.N. Yang Institute for Theoretical Physics, SUNY, Stony Brook, NY 11794-3840, USA}

\begin{abstract}
A variational ansatz for momentum eigenstates of translation invariant quantum spin chains is formulated. The matrix product state ansatz works directly in the thermodynamic limit and allows for an efficient implementation (cubic scaling in the bond dimension) of the variational principle. Unlike previous approaches, the ansatz includes topologically non-trivial states (kinks, domain walls) for systems with symmetry
breaking. The method is  benchmarked using the spin-\textonehalf\ XXZ antiferromagnet and the spin-1 Heisenberg antiferromagnet and we obtain surprisingly accurate results.
\end{abstract}

\maketitle

The density matrix renormalization group (DMRG) has proven to be the most successful variational method for strongly correlated quantum lattice systems in one spatial dimension.\cite{dmrg} The associated variational class is the class of matrix product states (MPS),\cite{mps} which has been generalized to higher dimensional systems\cite{peps} and can also be applied directly in the thermodynamic limit.\cite{itebd} It is now understood that the success of these ans\"atze can be attributed to the fact that they occupy that corner of Hilbert space that is characterized by an area scaling law of entanglement entropy, a property also satisfied by groundstates of gapped short-ranged quantum systems.\cite{arealaws} The same argument applies equally --- under some general constraints --- to the lowest excited states of such systems.\cite{arealawsexcitedstates}

Accurate information about the lowest-lying excited states is important to relate theoretical models to experimental measurements via spectral functions. Excited states appear as poles in these spectral functions, with corresponding residues given by the spectral weight. In dynamic DMRG methods,\cite{ddmrg} information about the spectrum of excited states is gathered from an approximate computation of the spectral function. The latest state-of-the-art algorithms first generate time-dependent correlation functions through a dynamic real-time evolution, after which highly accurate information about the spectrum can be extracted using spectral analysis.\cite{dynamic} However, this approach is limited by the fact that only reasonably short time scales are computationally accessible, due to the linear growth of entanglement under real-time evolution. This results in a broadening of the exact poles in spectral functions. Extracting high quality information requires a combination of working with a large bond dimension $D\approx \mathcal{O}(10^{3})$, linear prediction to extend the range of accessible time scales and complex statistical machinery to extract the precise position of the pole. However, since low-lying excited states also satisfy an area law for the scaling of entanglement entropy, it should be possible to construct a more direct and efficient approximation.

Nevertheless, MPS-inspired variational ans\"atze for excited states are rare. Most interesting is the case of translation-invariant states, where the Hamiltonian is block-diagonal in the different momentum sectors. Rommer and \"Ostlund proposed a Bloch-like ansatz that allowed them to get an early estimate of the Haldane gap in the spin-1 Heisenberg antiferromagnet,\cite{rommerostlund} by adding a virtual boundary operator $Q$ acting in the $D$-dimensional auxiliary space to the MPS approximation of the ground state and making a momentum superposition thereof. This is closely related to the general strategy of Bijl, Feynman, and Cohen, who act with the Fourier transform of a local physical operator $\operator{O}$ on the ground state to create excitations \cite{feynmanbijl}. This strategy is called the single-mode approximation in the context of spin systems,\cite{sma} and it has been applied to MPS in \cite{smamps}. Other ans\"atze include the projected entangled momentum states \cite{pems} and very recently the proposal by Pirvu \textit{et al.},\cite{pirvu} in which a momentum superposition is taken of the ground state MPS in which at a single site the set of matrices $A^{s}$ are replaced by the set of matrices $B^{s}$ that is variationally optimized. This ansatz contains and extends the Rommer and \"Ostlund ansatz ($B^{s}=Q A^{s}$) and the single-mode approximation ($B^{s}=\sum_{t}\braket{s|\operator{O}|t}A^{t}$). All of these proposals exploit translational invariance on a finite lattice with periodic boundary conditions, which unfortunately introduces finite-size effects and prevents them from reaching the computational efficiency [$\mathcal{O}(D^{3})$ with $D$ the bond dimension of the MPS] of indirect methods on systems with open boundary conditions.

This paper introduces a variational ansatz that allows us to describe excited states directly in the thermodynamic limit. This ansatz generalizes,\cite{pirvu} but differs by not relying on periodic boundary conditions, which is of key importance for the formulation of a computationally efficient implementation. This also allows for the possibility of topologically non-trivial excited states, which are very important in systems with symmetry breaking,\cite{fadeev} but for which few direct alternatives are available.

We now consider a one-dimensional lattice of $d$-level quantum systems described by  a local, translation-invariant Hamiltonian $\operator{H}=\sum_{n\in\mathbb{Z}} \operator{T}^{n} \operator{h} \operator{T}^{-n}$, with $\operator{T}$ the translation operator that shifts the lattice over a single site, and $\operator{h}$ an operator that acts nontrivially only on sites zero and one (we restrict to nearest-neighbor Hamiltonians for the sake of simplicity). We approximate translation-invariant ground states of such Hamiltonians with infinite size uniform MPS (uMPS), given by
\begin{displaymath}
\textstyle
\ket{\Psi(A)}= v_{\mathrm{L}}^{\dagger}\left(\prod_{n\in \mathbb{Z}} \sum_{s_{n}=1}^{d} A^{s_{n}}\right)v_{\mathrm{R}}|\mathbf{s}\rangle ,
\end{displaymath}
where $|\mathbf{s}\rangle \equiv \ket{\ldots s_{1}s_{2}\ldots}$, $A^{s}$ ($s=1, 2, \ldots, d$), constitute a set of $D\times D$ complex matrices acting on a $D$-level \emph{auxiliary} system, and $v_{\mathrm{L}}$ and $v_{\mathrm{R}}$ are two $D$-dimensional vectors living at $\pm \infty$. The MPS construction has a \emph{gauge invariance} under the gauge transform $A^{s}\mapsto G A^{s} G^{-1}$, with $G$ an invertible $D\times D$ matrix. While leaving the gauge unspecified, we do assume that the \emph{transfer matrix} $E^{A}_{A}= \sum_{s=1}^d A^{s}\otimes \bar{A}^{s}$ has precisely one eigenvalue $1$ with corresponding left and right eigenvectors $\rbra{l}$ and $\rket{r}$ of length $D^2$, to which we can associate $D\times D$ matrices $l$ and $r$, respectively, by reshaping them. These two matrices are Hermitian and positive and are assumed to be full rank. We choose the normalization $\rbraket{l|r}=\mathrm{Tr}(l r)=1$. In addition, we assume that all other eigenvalues of $E^{A}_{A}$ lie strictly within the unit circle, so  the spectral radius of $E^{A}_{A}-\rket{r}\rbra{l}$ is smaller than $1$ \cite{footnote1}. Under these conditions, the boundary vectors $v_{\mathrm{L}}$ and $v_{\mathrm{R}}$ do not feature in normalized expectation values of local operators.

Within the philosophy of Bijl, Feynman and Cohen, a typical elementary excitations of a local gapped Hamiltonian can be interpreted as a momentum superposition of a localized disturbance of the ground state. We therefore define a variational ansatz for  excitations as
\begin{displaymath}
\ket{\Phi_{\kappa}(B)}=\sum_{n\in \mathbb{Z}} \mathrm{e}^{\ic \kappa n}\operator{T}^{n} v_{\mathrm{L}}^{\dagger}\left(\cdots A^{s_{-1}} B^{s_{0}} \tilde{A}^{s_{1}}\cdots\right)v_{\mathrm{R}}|\mathbf{s}\rangle,
\end{displaymath}
where $A^{s}$ and $\tilde{A}^{s}$ represent the same (for a topologically trivial excitation) or different (for a topologically non-trivial excitation in the case of symmetry breaking) set of matrices for which $\ket{\Psi(A)}$ and $\ket{\Psi(\tilde{A})}$ are \emph{equally good} (\textit{i.e.}\ same energy) uMPS approximations of  ground states of $\operator{H}$ (\textit{e.g.}\ obtained using the imaginary time-dependent variational principle \cite{tdvpmps}). The state $\ket{\Phi_{\kappa}(B)}$ has momentum $\kappa \in [-\pi,\pi)$. The set of $D\times D$ matrices $B^{s}$ ($s=1,2,\ldots,d)$ contains the only variational parameters in our class; they are also denoted as a $D^{2}d$ vector $B$. All expectation values are quadratic in $B$, and we define
\begin{align*}
\braket{\Phi_{\kappa}(B)|\Phi_{\kappa'}(B')}&=2\pi\delta(\kappa-\kappa') B^{\dagger} \mathsf{N}_{\kappa} B',\\
\braket{\Phi_{\kappa}(B)|\operator{H}-H|\Phi_{\kappa'}(B')}&=2\pi\delta(\kappa-\kappa') B^{\dagger} \mathsf{H}_{\kappa} B'.
\end{align*}
For an infinite system size, the momentum eigenstates $\ket{\Phi_{\kappa}(B)}$ cannot be normalized to $1$ but rather satisfy a $\delta$ normalization. This $\delta$-function also appears in the expectation value of every translation invariant operator. The energy expectation value $\braket{\Phi_{\kappa}(B)|\operator{H}|\Phi_{\kappa'}(B')}$ has a contribution $H\braket{\Phi_{\kappa}(B)|\Phi_{\kappa'}(B')}$, where $H=\braket{\Psi(A)|\operator{H}|\Psi(A)}=|\mathbb{Z}| \braket{\Psi(A)|\operator{h}|\Psi(A)}$ is the diverging ground state energy, which was therefore subtracted in the definition of $\mathsf{H}^{(\Phi)}_{\kappa}$. The spectrum of excitation energies $\omega$ at momentum $\kappa$ can then be obtained by solving the $dD^{2}$-dimensional generalized eigenvalue problem $(\mathsf{H}^{(\Phi)}_{\kappa},\mathsf{N}^{(\Phi)}_{\kappa})$. Since our variational space is a linear subspace of the Hilbert space, this generalized eigenvalue system can be recognized as the Rayleigh-Ritz equation.

However, it can easily be seen that $\forall X \in\mathbb{C}^{D\times D}$, the choice $B^{s}=\mathrm{e}^{\ic \kappa} A^{s} X - X \tilde{A}^{s}$ results in $\ket{\Phi_{\kappa}(B)}=0$. For any $\kappa\neq 0$, or for $\kappa=0$ and $\ket{\Psi(A)}\neq \ket{\Psi(\tilde{A})}$, our linear parametrization $B$ has $D^{2}$ linearly independent \emph{zero modes} which can be eliminated by fixing a part of the variational parameters. For $\ket{\Psi(A)}=\ket{\Psi(\tilde{A})}$, there exists a gauge transformation $G$ such that $A^{s}=G \tilde{A}^{s}G^{-1}$ and one can see that $X=G$ leads to $B^{s}=0$ for $\kappa=0$. Hence, there are only $D^{2}-1$ linearly independent zero modes, but we can fix one additional variational parameter in $B$ by imposing $\braket{\Psi(A)|\Phi_{\kappa}(B)}=0$. This orthogonality constraint is automatically satisfied in all other cases. Hence, in all cases only $(d-1)D^{2}$ variational parameters remain. We prove elsewhere that this freedom in fixing some variational parameters is related to the gauge freedom in the original manifold of MPS, and that we can construct a linear representation $B(x)$ in terms of a $(d-1)D\times D$ matrix $x$ containing the free variational parameters such that $B(x)^{\dagger} \mathsf{N}_{\kappa}^{(\Phi)} B(y)=\mathrm{tr}[x^{\dagger} y]$ \cite{inpreparation}. In this study, we find the $dD\times(d-1) D$ matrix $V_{L}$ that contains an orthonormal basis for the null space of the $D\times dD$ matrix $L$ with entries $L_{\alpha,(\beta s)}=[(A^{s})^{\dagger}l^{1/2}]_{\alpha,\beta}$. Reshaping $V_{L}$ such that $[V_{L}^{s}]_{\alpha\beta}= [V_{L}]_{(\alpha s),\gamma}$ (for all $\alpha=1,\ldots,D$, $s=1,\ldots,d$, $\gamma=1,\ldots,(d-1)D$), the representation is given by $B^{s}(x)=l^{-1/2} V_{L}^{s} x \tilde{r}^{-1/2}$, with $l$ and $\tilde{r}$ the left and right eigenvector of $E^{A}_{A}$ and $E^{\tilde{A}}_{\tilde{A}}$. We then obtain (see \cite{inpreparation})
\begin{align*}
&B(x)^{\dagger} \mathsf{H}_{\kappa} B(y)= \big[\rbraket{l|H^{B(y)\tilde{A}}_{B(x) \tilde{A}}|\tilde{r}}+\rbraket{l|H^{AB(y)}_{AB(x)}|\tilde{r}}\\
&\ +\mathrm{e}^{+\ic\kappa}\rbraket{l|H^{AB(y)}_{B(x)\tilde{A}}|\tilde{r}}+\mathrm{e}^{-\ic\kappa}\rbraket{l|H^{B(y)\tilde{A}}_{AB(x)}|\tilde{r}}\\
&\ +\rbraket{l|E^{B(y)}_{B(x)}(1-E^{\tilde{A}}_{\tilde{A}})^{-1}H^{\tilde{A}\tilde{A}}_{\tilde{A}\tilde{A}}|\tilde{r}}\\
&\ +\rbraket{l|H^{AA}_{AA}(1-E^{A}_{A})^{-1}E^{B(y)}_{B(x)}|\tilde{r}}\\
&\ +\mathrm{e}^{+\ic\kappa}\rbraket{l|H^{AA}_{AA}(1-E^{A}_{A})^{-1}E^{A}_{B(x)}(1-\mathrm{e}^{+\ic\kappa}E^{A}_{\tilde{A}})^{-1}E^{B(y)}_{\tilde{A}}|\tilde{r}}\\
&\ +\mathrm{e}^{-\ic\kappa}\rbraket{l|H^{AA}_{AA}(1-E^{A}_{A})^{-1}E^{B(y)}_{A}(1-\mathrm{e}^{-\ic\kappa}E^{\tilde{A}}_{A})^{-1}E^{\tilde{A}}_{B(x)}|\tilde{r}}\\
&\ +\mathrm{e}^{+\ic\kappa}\rbraket{l|H^{AA}_{AB(x)}(1-\mathrm{e}^{+\ic\kappa}E^{A}_{\tilde{A}})^{-1}E^{B(y)}_{\tilde{A}}|\tilde{r}}\\
&\ +\mathrm{e}^{-\ic\kappa}\rbraket{l|H^{AB(y)}_{AA}(1-\mathrm{e}^{-\ic\kappa}E^{\tilde{A}}_{A})^{-1}E^{\tilde{A}}_{B(x)}|\tilde{r}}\\
&\ +\mathrm{e}^{+2\ic\kappa}\rbraket{l|H^{AA}_{B(x)\tilde{A}}(1-\mathrm{e}^{+\ic\kappa}E^{A}_{\tilde{A}})^{-1}E^{B(y)}_{\tilde{A}}|\tilde{r}}\\
&\ +\mathrm{e}^{-2\ic\kappa}\rbraket{l|H^{B(y)\tilde{A}}_{AA}(1-\mathrm{e}^{-\ic\kappa}E^{\tilde{A}}_{A})^{-1}E^{\tilde{A}}_{B(x)}|\tilde{r}}\big],
\end{align*}
where $E^{A}_{B}=\sum_{s=1}^d A^{s}\otimes \overline{B}^{s}$ and $H^{AB}_{CD}=\sum_{s,t,u,v=1}^d \braket{s,t|\operator{h}|u,v}(A^{u}B^{v}) \otimes(\overline{C}^{s}\overline{D}^{t})$. When $A\neq \tilde{A}$, which does not necessarily imply $\ket{\Psi(A)}\neq\ket{\Psi(\tilde{A})}$, we can substitute $\tilde{A}\leftarrow \mathrm{e}^{\ic \varphi}\tilde{A}$ in order to obtain $\ket{\Phi_{\kappa}(B)}=\ket{\Phi_{\kappa-\varphi}(B)}$ up to an infinite phase, which seems to indicate that the momentum label is completely arbitrary. This is an artifact of not having momentum in a system with open boundary conditions. It does not appear when $\tilde{A}=A$. This inconsistency can be solved by fixing $\varphi$ such that the dominant eigenvalue (largest in magnitude) of $E^{\tilde{A}}_{A}$ is positive. When $\ket{\Psi(A)}=\ket{\Psi(\tilde{A})}$ up to phase, the dominant eigenvalue of $E^{\tilde{A}}_{A}$ is $1$. By assumption, the dominant eigenvalue of $E^{A}_{A}$ and $E^{\tilde{A}}_{\tilde{A}}$ are always $1$. All inverses of $(1-E^{A}_{A})$, $(1-E^{\tilde{A}}_{\tilde{A}})$, $(1-\mathrm{e}^{\ic \kappa}E^{A}_{\tilde{A}})$ and $(1-\mathrm{e}^{-\ic \kappa}E^{\tilde{A}}_{A})$ should be read as `pseudo-inverses' \cite{footnote2} that act as zero in the eigenspace corresponding to the eigenvalue one of the transfer operator. In the case of symmetry breaking with $\ket{\Psi(A)}\neq \ket{\Psi(\tilde{A})}$, the spectral radius $\rho(E^{\tilde{A}}_{A}) < 1$ and the expressions $(1-\mathrm{e}^{\ic \kappa}E^{A}_{\tilde{A}})^{-1}$ and $(1-\mathrm{e}^{-\ic \kappa}E^{\tilde{A}}_{A})^{-1}$ denote the full inverses. If $A$ and $\tilde{A}$ satisfy the properties that were outlined before, it is straightforward to prove that the corresponding uMPS approximate ground states with \emph{maximal symmetry breaking}, \textit{i.e.} they yield extremal values for the expectation value of the order parameter associated with the symmetry breaking.

Excitation energies can thus be found from diagonalizing the effective $(d-1)D^{2} \times (d-1) D^{2}$ Hamiltonian defined with respect to entries of $x$ and $y$ in $B(x)^{\dagger}\mathsf{H}_{\kappa} B(y)$, since the effective norm matrix is now the unit matrix. As shown in the results below, we are often interested in the lowest excitation energies. Using an iterative method for the different (pseudo)-inversions, the action of the effective Hamiltonian can be implemented as an $\mathcal{O}(D^{3})$ operation and can be combined with a sparse eigensolver.

\begin{figure}
\begin{minipage}{\columnwidth}
\includegraphics[width=\columnwidth]{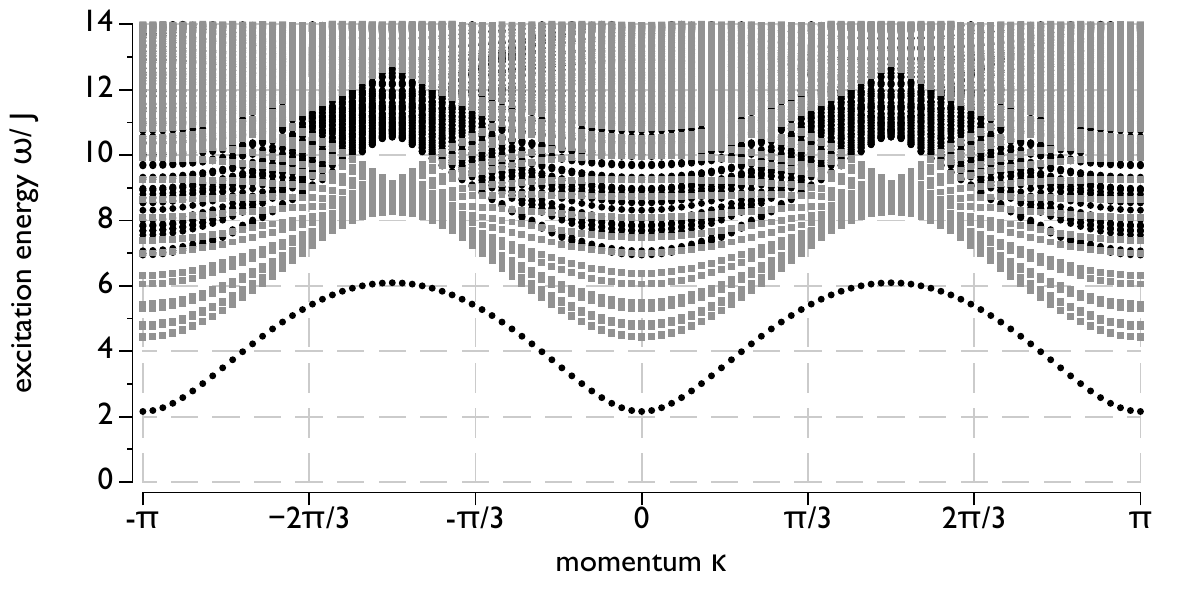}
\caption{Spectrum of the lowest lying excitations of the spin-\textonehalf\ XXZ antiferromagnet with anisotropy parameter $\Delta=4$ at $D=33$. Black circles indicate topologically non-trivial excitations, gray squares indicate topologically trivial excitations.}
\label{fig:xxzspectrum}
\end{minipage}
\begin{minipage}{\columnwidth}
\includegraphics[width=\columnwidth]{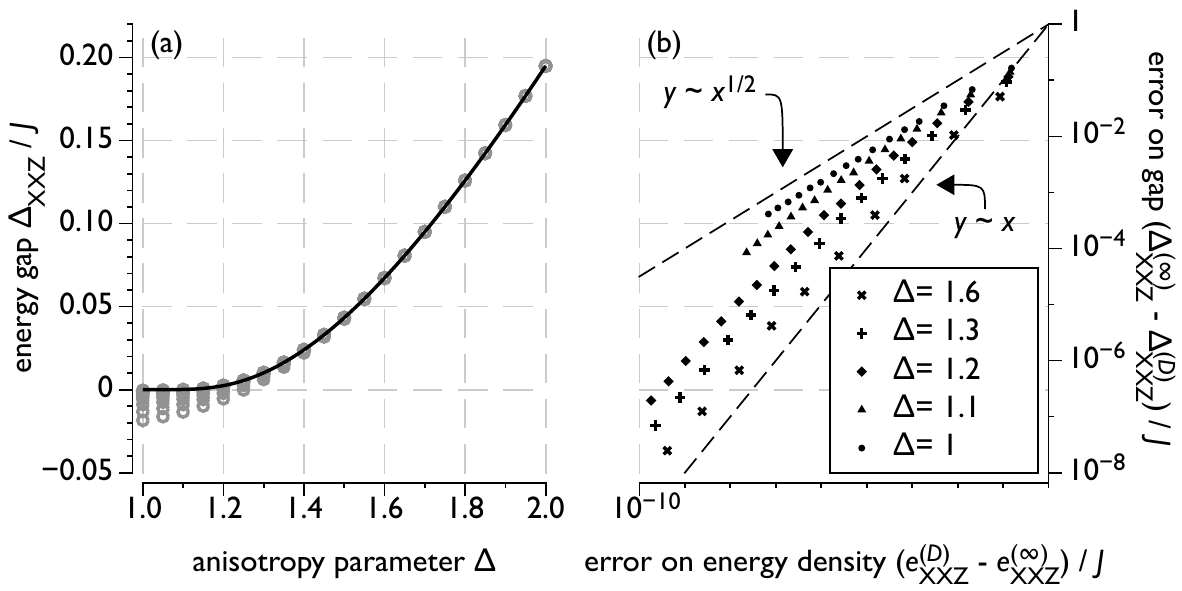}
\caption{(a) Simulation results for $\Delta_{\text{XXZ}}^{(D)}$ as a function of the anisotropy $\Delta$ for various values of $D$ ranging from $10$ to $400$ (gray circles), as well as the exact result with the Bethe ansatz (black line). (b) Absolute error on the energy gap versus absolute error on the energy density for various values of $\Delta$ and $D$.}
\label{fig:xxzgap}
\end{minipage}
\end{figure}

We now illustrate the power of our variational ansatz for excited states using some benchmark problems. The first Hamiltonian under consideration is the spin-\textonehalf\ XXZ antiferromagnet in the symmetry-breaking phase $\Delta>1$. The ground states of maximal symmetry breaking are antiferromagnetic and also break translational invariance. We therefore perform a spin-flip ($\sigma^{x}$) on every second site, in order to obtain
\begin{displaymath}
\operator{H}_{\text{XXZ}}=J\sum_{n\in \mathbb{Z}}\sigma^{x}_{n}\sigma^{x}_{n+1}-\sigma^{y}_{n}\sigma^{y}_{n+1}-\Delta \sigma^{z}_{n}\sigma^{z}_{n+1}.
\end{displaymath}
The gap closes at the critical point $\Delta=1$, where we obtain the spin-\textonehalf\ Heisenberg antiferromagnet. Fig.~\ref{fig:xxzspectrum} displays the full spectrum of excited states obtained with our ansatz at $D=33$ (full diagonalization becomes computationally demanding for much larger values of $D$) for $\Delta=4$, resulting in $(d-1)D^{2}=1089$ topologically trivial excitations ($\tilde{A}^{s}=A^{s}$) and $1089$ topologically non-trivial excitations ($\tilde{A}^{s}=\sum_{t}\braket{s|\sigma^{x}|t}A^{t}$). As pointed out in \cite{fadeev}, the elementary particle excitations in the symmetry-broken phase are topologically non-trivial kinks, and all topologically trivial excitations are compound states containing an even number of kinks. Not only do we recover the elementary kink, we also obtain a large set of points that fall within the two-particle (topologically trivial) and three-particle (topologically non-trivial) continuum. This happens because our ansatz contains a single perturbation that can spread out over a region of $\mathcal{O}(\log_{d} D)$ sites. The two-particle states are states with fixed total momentum $\kappa_{1}+\kappa_{2} \mod 2\pi = \kappa$, but which consist of a superposition of relative momenta $\triangle \kappa=\kappa_{2}-\kappa_{1}$ so as to confine the two particles into the region allowed by the ansatz. 

We can assess the accuracy of our approach as a function of $D$ by comparing the energy gap $\Delta_{\text{XXZ}}^{(D)}$ (lowest excitation energy at $\kappa=0$ or $\kappa=\pi$) with the exact value $\Delta_{\text{XXZ}}^{(\infty)}$. Because this gap belongs to a topologically non-trivial excitation that only comes in pairs on lattices with periodic boundary conditions, the value of the energy gap calculated in \cite{cloizeaux} using the Bethe ansatz on a lattice with periodic boundary conditions is twice the exact value. As illustrated in Fig.~\ref{fig:xxzgap}, we can even obtain highly accurate values of the energy gap very close to the critical point, by going to larger values of the bond dimension $D$ using a sparse implementation. Note that errors on the elementary excitation are negative. This violation of the variational principle is caused by subtracting an estimate of the ground state energy $H$ that is too large. Also note that the error on the gap scales as the square root of the error on the energy (density) for low values of $D$, but is proportional to this error for larger $D$, except at the critical point.

\begin{figure}
\includegraphics[width=\columnwidth]{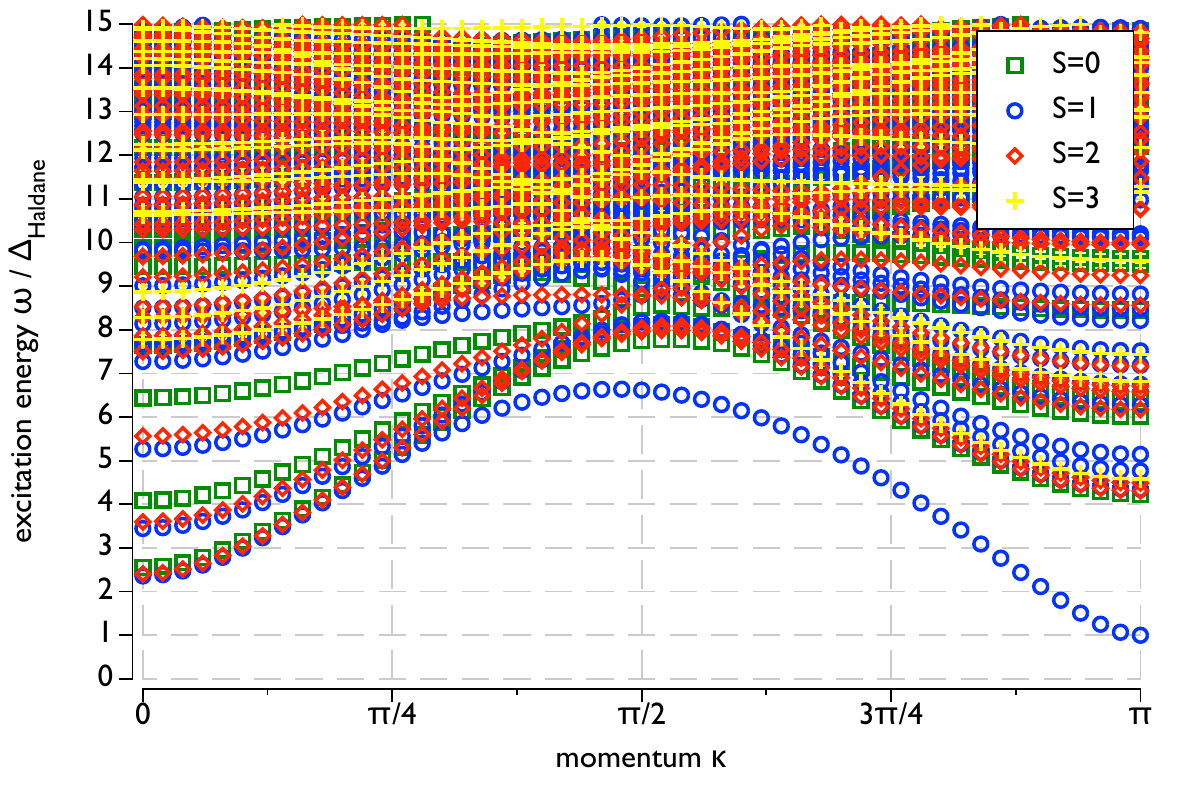}
\caption{Spectrum of the lowest lying excitations of the spin-1 Heisenberg antiferromagnet at $D=30$, labeled by their spin $S$ degeneracy (color online).}
\label{fig:heisenbergspectrum}
\end{figure}

Secondly, we study the spin-1 Heisenberg antiferromagnet, which has a  translation-invariant ground state in a symmetry-protected topological phase and is characterized by the presence of the Haldane gap.\cite{haldanegap} There is a long history of numerical estimates of the Haldane gap using a variety of methods\cite{dynamic,rommerostlund,haldangegapestimate}. Fig.~\ref{fig:heisenbergspectrum} shows the spectrum of topologically trivial excitations of the spin-1 Heisenberg antiferromagnet, obtained using our ansatz for $D=30$, where excitation energies are colored according to their degeneracy (which was obtained without using the symmetry explicitly). As expected, the elementary excitation around momentum $\kappa=\pi$ is the $S=1$ magnon excitation, and the corresponding energy at $\kappa=\pi$ is the Haldane gap. This spectrum is in agreement with a previous proposal that was obtained using a dynamic DMRG simulation on a finite chain of up to 400 sites with values for the bond dimension up to $D=2000$ (second reference in \cite{dynamic}). The most accurate results for the Haldane gap are obtained using ground state DMRG on finite lattices in the second reference of \cite{dynamic} ($\Delta=0.41047925(4)$ on a lattice of $400$ sites with $D=500$) and the last reference of \cite{haldanegapestimate} ($\Delta=0.4104792485(4)$ on a lattice of $2048$ sites with $D$ up to $2700$). With modest computational resources (solving a single eigenvalue problem iteratively) we obtain similar results for various values of $D$ up to $D=208$, which were chosen so that all Schmidt values with a certain degeneracy (given by half-integral spin representations) are present, but without explicitly using the $\mathsf{SU}(2)$ symmetry. We can even improve the estimate for the Haldane gap by two more significant digits from a scaling analysis in $D$: $\Delta= 0.410479248463^{+6\times 10^{-12}}_{-3\times 10^{-12}}$. Unlike in the second reference in \cite{dynamic}, where the rest of the dispersion relation was much less accurate than the gap, we can now expect roughly the same accuracy for all points of the dispersion relation where the elementary magnon exists. Around $\kappa=\pi/4$, the elementary magnon excitation is absorbed into the two-magnon continuum and becomes unstable against decay into two magnons.

We have presented a variational algorithm, based on the matrix product state formalism, to determine topologically trivial and non-trivial excited states of one-dimensional quantum lattices, directly in the thermodynamic limit. We envisage that this set of excited states can also be used to accurately determine spectral functions, as requested in \cite{spectralfunctionsnew}. In addition, we expect that our proposal can be extended to the setting of two-dimensional lattice systems, by replacing a single tensor in the network of projected entangled-pair states\cite{peps}, or to the setting of quantum field theories, by building on the continuous MPS proposal\cite{cmps}.

\acknowledgements{J.H.,\ H.V.\ and D.J.W.\ would like to thank R.\ Bertlmann and F.V.\ for inviting them to the University of Vienna, where this research was initiated. D.J.W. is also grateful to J.H.\ and H.V.\ for their kind hospitality at the University of Ghent. Research supported by the Research Foundation Flanders (J.H.), the FWF doctoral program Complex Quantum Systems (W1210) (B.P.), the Science and Technology Facilities Council (D.J.W.), the DFG (FOR 635 and SFB 631) (J.I.C.), the EU Strep project QUEVADIS, the ERC grant QUERG and the FWF SFB grants FoQuS and ViCoM.


\begin{thebibliography}{99}
\bibitem{footnote1}
States satisfying this condition are also known as pure finitely correlated states.
\bibitem{footnote2}
For $\kappa \neq 0$, the operators $(1-\mathrm{e}^{\ic \kappa}E^{A}_{\tilde{A}})$ and $(1-\mathrm{e}^{-\ic \kappa}E^{\tilde{A}}_{A})$ are not really singular and could also be inverted fully. This will lead to the same result due to our representation of $B$.
\end{thebibliography}
\end{document}